# Sustainability and Reproducibility via Containerized Computing*

R. Nagler,[#] D.L. Bruhwiler,[#,$] P. Moeller[%] and S.D. Webb[#]

[#]RadiaSoft LLC, Boulder, CO 80304, USA

[$]RadiaBeam Technologies LLC, Santa Monica, CA 90404, USA

[%]Bivio Software Inc., Boulder, CO 80303, USA

*Abstract*

Recent developments in the commercial open source community have catalysed the use of Linux containers for scalable deployment of web-based applications to the cloud. Scientific software can be containerized with dependencies, configuration files, post-processing tools and even simulation results, referred to as containerized computing. This new approach promises to significantly improve sustainability, productivity and reproducibility. We present our experiences, technology, and future plans for open source containerization of software used to model particle and radiation beams. Vagrant is central to our approach, using Docker for cloud deployment and VirtualBox virtual machines for deployment to Mac OS and Windows computers. Our technology enables seamless switching between the desktop and the cloud to simplify simulation development and execution.

## Introduction

Reproducibility is an essential requirement for scientific advancement, yet it is difficult and time-consuming to achieve for large-scale simulations of particle accelerators and other physical systems. Open source software and open science principles are important for reproducibility, but significant technical difficulties can still prevent success. For example, detailed reproduction of previous work may require access to the same version of an application and multiple dependencies, the ability to build and install with the same compiler and flags, access to the original input files or other configuration details, and use of the same or similar visualization tools.

Recent developments in open source software for application containers (e.g. Docker [1], Vagrant [2], VirtualBox [3]) have made it possible to containerize scientific simulations for archival and collaboration, which immediately enables other scientists to reproduce and extend previous work. The implications of containerization for high performance computing is beginning to receive attention [4,5]. We are developing complete use cases of containerized computing for particle accelerator and radiation codes.

The software we plan to support in the near term includes Elegant [6], SRW [7-10], Synergia [11], WARP [12-15] and SHADOW [16,17] Cross-platform installers are being developed for Mac OS and Windows, enabling single-click access to multiple scientific codes within a Linux virtual machine, including all necessary input files, scripts and post-processing tools. Innovative use of tools like pyenv [18] are key to the successful management and deployment of multiple scientific codes.

Our vision is to bring scientific cloud computing to the accelerator technology and radiation source communities. We believe scientific software must be open source, so all of our cloud computing infrastructure will be available on GitHub. Our vision has three primary components: a) containerized computing; b) the browser is the UI; and c) seamless support of both desktop and cloud computing, both web-based and command line interfaces.

## Containerized Computing

Some scientific software development teams have expended significant effort to achieve cross-platform execution on Linux, Mac OS and Windows; however, this is time consuming, expensive, and still requires sophistication on the part of the user to correctly install and use such codes. Dependencies often include a specific version of Python and other libraries, which significantly complicate installation and can sometimes clash on any OS with previously installed software. Even different flavors of Linux can cause serious pain for users trying to install a scientific code, especially when using a cluster where many of the required dependencies are not installed, or the system installation is the wrong release.

Using Docker on Linux, it is possible to create a file that contains a scientific code or codes, plus all required tools and dependencies, which can then be copied to any Linux server or cluster and rapidly activated. A user can ssh into the container, if necessary, or the software can be accessed remotely through a web-based UI. This removes the pain of software installation on Linux, and it enables cloud-based scientific computing by providing on-demand access via local cluster, supercomputer or commercial service.

There is no runtime overhead associated with the use of Dockerized software on Linux. On Windows and Mac OS, the containerized software must run inside a Linux virtual machine; however, Vagrant and headless invocation of VirtualBox make this possible with a minimum of runtime overhead. We are developing open source single-click installers to hide these complications from the user.

Containerization allows scientists to archive their entire simulation environment in the cloud, then return to their work weeks or months later. More importantly, such an archive could be published together with a refereed journal article, so that readers are able to interactively explore and reproduce the published simulation results, using the same version(s) of the code and its dependencies, already compiled in exactly the same way. Collaborators will also benefit from this ability to share a complete simulation environment "in a box". Commercial companies [19,20] are beginning to offer services along these lines to other communities.

## The Browser Is The UI

Our vision is that the web browser will become the ubiquitous user interface (UI) for scientific computing. This ambitious goal has become viable very recently, due to the emergence of powerful, standardized technologies, including HTML5 [21], CSS [22], JavaScript [23,24] and scalable vector graphics (SVG) [25]. There are numerous technologies for scientific visualization, which build on these standards.

## Seamless Legacy

We respect the existing workflows of computational scientists, so our vision includes support for both command line and web based UIs. Also, any code we provide via the cloud will also be available for use on desktop and laptop computers.

Many scientific codes use cross-platform UIs such as Qt [26] and Matplotlib [27]; however, such applications are not suitable for cloud computing. While these applications can be run remotely over X11, it requires the end-user have an X11 server installed on their desktop and to have ssh access to the remote computer. In order to accomplish our goal of seamless legacy integration, we are implementing X11 in the browser. Proof-of-concept Javascript X11 server implementations are available (XPlain [29] and GateOne[30]). We will expand on this work to build instantaneous access to legacy scientific codes.

## Acknowledgments

This work is supported by the US DOE Office of Science, Office of Basic Energy Sciences through Grant No.'s DE-SC0006284 and DE-SC0011237, and by the Office of High Energy Physics through Grant No.'s DE-SC0011340 and DE-SC0013855.